\documentclass[12pt]{article}
\usepackage[cp1251]{inputenc}
\usepackage{amsmath}
\usepackage{amsfonts}
\usepackage{amssymb}
\def\pa{\partial}                       
\def\beq{\begin{eqnarray}}    
\def\eeq{\end{eqnarray}}      

\begin{document}
\date{}

\begin{center}
{\Large\textbf{Generalized canonical approach to deformation problem
in gauge theories}}

\vspace{18mm}

{\large I.L. Buchbinder$^{(a,b,c)}\footnote{E-mail:
joseph@tspu.edu.ru}$\;,
P.M. Lavrov$^{(a,c)} \footnote{E-mail:
lavrov@tspu.edu.ru}$,\;
}

\vspace{8mm}

\noindent  ${{}^{(a)}}
${\em
Center of Theoretical Physics, \\
Tomsk State Pedagogical University,\\
Kievskaya St.\ 60, 634061 Tomsk, Russia}

\noindent ${{}^{(b)}}
${\em
Bogoliubov Laboratory of Theoretical Physics,
JINR,\\
141980 Dubna, Moscow region, Russia}

\noindent  ${{}^{(c)}}
${\em
National Research Tomsk State  University,\\
Lenin Av.\ 36, 634050 Tomsk, Russia}

\vspace{20mm}

\begin{abstract}
\noindent
We develop a general approach to constructing a deformation that
describes the mapping of
any dynamical system with irreducible first-class constraints in the
phase space into another
dynamical system with first-class constraints.  It is shown that such
the deformation problem can be
efficiently embedded  in the Batalin-Fradkin-Vilkovisky (BFV) formalism by
using  (super)canonical transformations
which leave the basic equations of the formalism invariant. It is proved that
the generating function of (super)canonical transformations   is determined
by a single function which depends only on coordinates of initial
dynamical system.
To illustrate the developed approach, we have constructed a non-local
deformation of
the Abelian gauge theory into a non-local non-Abelian gauge theory
whose local sector
coincides with the standard Yang-Mills theory.

\end{abstract}

\end{center}

\vfill

\\

\newpage

\section{Introduction}
\noindent

The gauge theories are the key elements of the Standard Model of
Fundamental Interactions and its generalizations. Therefore,
studying the structure of general gauge theories can in principle
ensure the possibilities to formulate the new concrete gauge
theories that can be useful for constructing the new models of
fundamental interactions beyond the standard model. Recently, we
have developed an approach to generate the new gauge theories on the
base of known gauge theories \cite{BL-1}, \cite{BL-2}, \cite{L}  and
describe their quantum aspects. The approach was formulated within
the framework of the Batalin-Vilkovisky (BV) - formalism of
covariant quantization of general gauge systems \cite{BV},
\cite{BV1}.

In this paper we study a problem of generating the new gauge
theories on the base of known gauge theories on the other hand using
the Batalin-Fradkin-Vilkovisky (BFV) - formalism of canonical
quantization of general gauge systems \cite{BFV1}, \cite{BFV2},
\cite{BFV3} (see also the reviews \cite{H1}, \cite{BF1}, \cite{BF2}
and the book \cite{HT}). For some recent developments of this
formalism see e.g. \cite{BLavT}, \cite{BLav1}, \cite{BLav2} and the
references therein. The BFV-formalism, based on the fundamental
principle of the BRST symmetry \cite{brs1}, \cite{t}, is the powerful
quantization method for arbitrary dynamical systems with constraints
in phase space. The central object of the BFV-formalism is a
nilpotent BRST-BFV charge encoding the gauge structure in extended
phase space involving the ghost coordinates and momenta. In these
terms, a gauge invariant deformation of gauge theory means
(super)canonical transformation of extended phase space preserving
the nilpotency  of the BRST-BFV charge. We prove that such a
deformation can be described in general form in terms of generating
function which is defined by one arbitrary function of coordinates
of the initial dynamical system.

The paper is organized as follows. In section 2, we consider the
classical aspects of the BFV-formalism for dynamical systems with
irreducible first-class constraints. Section 3 is devoted to
formulating the problem of gauge invariant deformation in extended
phase space for gauge theories with first-class constraints and
discussing a way of a possible solution to this problem. In section
4, we describe the exact solution of the deformation problem in
closed form using the (super)canonical transformation in the initial
phase space. The generating function of such a transformation is
found in the explicit form. It should be noted that no special
properties of the above (super)canonical transformation are assumed
in advance. In particular, this transformation is not expected to be
necessarily local, however it may have a local sector corresponding
some new local gauge theory. In section 5, we apply the obtained
results to Abelian gauge theory and construct the special canonical
transformation such that a local part of the transformed gauge
theory coincides with non-Abelian Yang-Mills theory. Section 6 is a
summary of the results.

In this paper we systematically use the DeWitt's condensed notations \cite{DeWitt}
and apply the symbols $\varepsilon(A)$ for the Grassmann parity and
${\rm gh}(A)$ for the ghost number, respectively. The right and left
functional derivatives are marked by special symbols $"\leftarrow"$
and $"\overrightarrow{}"$ respectively.

\section{Classical aspects of the BFV- formalism }
\noindent

Our starting point is  a dynamical system described by a Hamiltonian
$H_0=H_0(q,p)$ in the phase space of canonically conjugate variables
$q^i$ and $p_i$
\beq
\label{qp}
\{q^i,p_j\}=\delta^i_{\;\!j}\;,
\quad \varepsilon(q^i)=\varepsilon(p_i)=\varepsilon_i,
\quad {\rm gh}(q^i)={\rm gh}(p_i)=0,\quad i=1,2,...,n ,
\eeq
and the set of
irreducible first-class constraints
\beq
\label{T}
T_{\alpha}=T_{\alpha}(q,p),\quad
\varepsilon(T_{\alpha})=\varepsilon_{\alpha},\quad \alpha=1,2,...,m
,
\eeq
which satisfy the involution relations,
\beq
\label{invTH}
\{T_{\alpha},T_{\beta}\}=T_{\gamma}U^{\gamma}_{\alpha\beta},\qquad
\{H_0,T_{\alpha}\}=T_{\beta} V^{\beta}_{\alpha}.
\eeq
The structure coefficients $U^{\gamma}_{\alpha\beta}$ and $V^{\beta}_{\alpha}$ are
in general functions of canonical variables $(q,p)$ with the
properties
$\varepsilon(U^{\gamma}_{\alpha\beta})=\varepsilon_{\alpha}+\varepsilon_{\beta}+
\varepsilon_{\gamma}$ and
$\varepsilon(V^{\beta}_{\alpha})=\varepsilon_{\alpha}+\varepsilon_{\beta}$.
The symbol $\{,\}$ means the Poisson superbracket defined for any
set of variables $(Q^A,P_A)$,
$\varepsilon(Q^A)=\varepsilon(P_A)=\varepsilon_A$ and any two
quantities $F=F(Q,P)$ and $G=G(Q,P)$ by the rule
\beq
\{F,G\}=F\big(\overleftarrow{\pa}_{\!Q^A}\;\overrightarrow{\pa}_{\!P_A}-
\overleftarrow{\pa}_{\!P_A}\;\overrightarrow{\pa}_{\!Q^A}\big)G ,
\eeq
and obeyed  the well-known algebraic properties for the superbracket
\cite{BFV1}.

Description of a given dynamical system within the framework of
BFV-formalism on classical level involves introduction of additional
degrees of freedom. For theories with the irreducible first-class
constraints (\ref{T}) they include the set of variables
\beq
(C^{\alpha}, {\cal P}_{\alpha}),\quad
\varepsilon(C^{\alpha})=\varepsilon({\cal
P}_{\alpha})=\varepsilon_{\alpha}+1, \quad {\rm
gh}(C^{\alpha})=-{\rm gh}({\cal P}_{\alpha})=1,\quad \alpha=1,2,...,
m.
\eeq
With the help of these variables one introduces the minimal
extended phase space of the BFV-formalism with the coordinates
\beq
\label{mps}
(Q^A,P_A)=(q^i,p_i, C^{\alpha}, {\cal P}_{\alpha}).
\eeq
On the
minimal extended phase space (\ref{mps}), the BRST-BFV charge,
$\Omega=\Omega(Q,P),\; \varepsilon(\Omega)=1$, and the generalized
Hamiltonian, ${\cal H}={\cal H}(Q,P), \;\varepsilon({\cal H})=0 $,
are defined as solutions to the equations
\beq \label{beOH}
\{\Omega,\Omega\} = 0, \quad \{\mathcal{H}, \Omega\}=0,
\eeq
satisfying  the boundary conditions
\beq
\label{bcOH}
\Omega\overleftarrow{\pa}_{\!C^{\alpha}}\big|_{C=0}=T_{\alpha},\quad
\mathcal{H}\big|_{C=0}=H_0.
\eeq
Solutions to the basic equations
(\ref{beOH}) with boundary conditions (\ref{bcOH}) always exist in
form of formal series with respect to ghost variables $C^{\alpha}$
\cite{BFV1}, \cite{BFV2}. In the lowest order in the ghosts one
gets
\beq
\label{sOH}
\Omega=T_{\alpha}C^{\alpha}+ \frac{1}{2}{\cal
P}_{\gamma}U^{\gamma}_{\alpha\beta}
C^{\beta}C^{\alpha}(-1)^{\varepsilon_{\alpha}}+\cdots, \qquad
\mathcal{H}=H_0+{\cal P}_{\beta}V^{\beta}_{\alpha}C^{\alpha}+\cdots .
\eeq
In the case when structure coefficients are constants, the
first two summands in (\ref{sOH}) present exact solutions to the
basic equations (\ref{beOH}).

In terms of basic quantities $\Omega$ and ${\cal H}$ of the BFV-formalism,
the quantization procedure is formulated. Here we do not concern the quantum aspects
of the BFV-formalism and concentrate only on classical ones having in mind the deformation.

\section{Deformation problem within the framework of canonical formalism}
\noindent
The deformation problem for theories with gauge freedom in Lagrangian formalism
is formulated in Ref. \cite{BH} (see also \cite{H}, \cite{D}).
In this section, we are going to
discuss aspects of this problem in the context of dynamical
systems with constraints in phase space.

We consider the dynamical system described in the previous section.
By deformation we will understand mapping characterized by the
deformation parameter $g$, transforming the initial quantities $H_0,
T_{\alpha},\;\alpha =1,2,...,m$ onto the deformed quantities
$\widetilde{H}_0, \widetilde{T}_{\alpha},\;\alpha =1,2,...,m$
defined in the same phase space, so that the following equations
hold
\beq
\{\widetilde{H}_0, \widetilde{T}_{\alpha}\}=
\widetilde{T}_{\beta}\widetilde{V}^{\beta}_{\;\!\alpha},\quad
\{\widetilde{T}_{\alpha},\widetilde{T}_{\beta}\}=
\widetilde{T}_{\gamma}\widetilde{U}^{\gamma}_{\;\;\alpha\beta},
\eeq
supported by the boundary conditions
\beq
\widetilde{H}_0\big|_{g=0}=H_0,\quad
\widetilde{T}_{\alpha}\big|_{g=0}=T_{\alpha},\quad
\widetilde{U}^{\gamma}_{\;\;\alpha\beta}\big|_{g=0}=U^{\gamma}_{\;\;\alpha\beta},\quad
\widetilde{V}^{\beta}_{\;\!\alpha}\big|_{g=0}=V^{\beta}_{\;\!\alpha}.
\eeq

In principle, there exist two ways to solve the deformation problem
under consideration. The first one can be based on the solution to
the deformation procedure in Lagrangian formalism \cite{BH}.
Following this way, we may propose to find solutions  to the
problem in Hamiltonian formalism in the form of series expansion
in parameter $g$,
\beq
&&\widetilde{H}_0=H_0+gH_{(1)}+g^2H_{(2)}+\cdots ,\qquad\qquad
\widetilde{T}_{\alpha}=T_{\alpha}+gT^{(1)}_{\alpha}+g^2T^{(2)}_{\alpha}+\cdots , \\
&&
\widetilde{U}^{\gamma}_{\;\;\alpha\beta}=U^{\gamma}_{\;\;\alpha\beta}
+gU^{(1)\gamma}_{\;\;\alpha\beta}+g^2U^{(2)\gamma}_{\;\;\alpha\beta}+\cdots,\qquad\!\!
\widetilde{V}^{\beta}_{\;\!\alpha}=
V^{\beta}_{\;\!\alpha}+gV^{(1)\beta}_{\;\!\alpha}+g^2V^{(2)\beta}_{\;\!\alpha}+\cdots.
\eeq
It yields the infinite system of equations,
\beq
\label{14}
&&\{H_0,
T^{(1)}_{\alpha}\}+\{H_{(1)}, T^{\alpha}\}=
T_{\beta}V^{(1)\beta}_{\;\!\alpha}+T^{(1)}_{\beta}V^{\beta}_{\;\!\alpha},\\
\label{15}
&&\{H_0, T^{(2)}_{\alpha}\}+\{H_{(1)}, T^{(1)}_{\alpha}\}+\{H_{(2)}, T^{\alpha}\}=
T_{\beta}V^{(2)\beta}_{\;\!\alpha}+
T^{(1)}_{\beta}V^{(1)\beta}_{\;\!\alpha}+T^{(2)}_{\beta}V^{\beta}_{\;\!\alpha},\\
\label{16}
&&\{T_{\alpha},T^{(1)}_{\beta}\}+\{T^{(1)}_{\alpha},T_{\beta}\}
=T_{\gamma}U^{(1)\gamma}_{\;\;\alpha\beta}+
T^{(1)}_{\gamma}U^{\gamma}_{\;\;\alpha\beta},\\
\label{17}
&&\{T_{\alpha},T^{(2)}_{\beta}\}+\{T^{(1)}_{\alpha},T^{(1)}_{\beta}\}+
\{T^{(2)}_{\alpha},T_{\beta}\}
=T_{\gamma}U^{(2)\gamma}_{\;\;\alpha\beta}+T^{(1)}_{\gamma}U^{(1)\gamma}_{\;\;\alpha\beta}+
T^{(2)}_{\gamma}U^{\gamma}_{\;\;\alpha\beta},
\eeq
and so on. Such an infinite system of equations is typical for deformation
procedure constructed in terms of infinite series. The
solution to this system can be described schematically, for example, as follows.
The first step in the solution begins with equations (\ref{14}) and (\ref{16}).
With known initial quantities $H_0$, $T^{\alpha}$, $V^{\beta}_{\;\!\alpha}$ and
$U^{\gamma}_{\;\;\alpha\beta}$ one has to find solutions to the equations
(\ref{14}) and (\ref{16}) in linear approximation with respect to the deformation parameter
$g$ for $H_{(1)}$, $T^{(1)}_{\alpha}$, $V^{(1)\beta}_{\;\!\alpha}$ and
$U^{(1)\gamma}_{\;\;\alpha\beta}$ using the cohomological methods \cite{BH,H}.
Then,  similarly, it needs to find solutions
to the equations (\ref{15}) and (\ref{17}) for quantities
$H_{(2)}$, $T^{(2)}_{\alpha}$, $V^{(2)\beta}_{\;\!\alpha}$ and
$U^{(2)\gamma}_{\;\;\alpha\beta}$ up to the second order in $g$. In principle,
the procedure must be continued up to infinity.
In general, solutions to these equations seem to be quite a complicated task.
\footnote{We are not going to discuss here the solutions of the above equations.
We have presented these equations solely to emphasize the difficulties of the
deformation problem at standard approach in terms of infinite series. New way
to study the deformation procedure that does not require series expansion is
considered in the next section.}

We also can attack the deformation problem considering the extended
phase space and applying the BFV-formalism. In this case, the
BRST-BFV charge $\widetilde{\Omega}$ and the generalized Hamiltonian
$\widetilde{{\cal H}}$ satisfy the generating equations
\beq
\label{baseq}
\{\widetilde{\Omega},\widetilde{\Omega}\}=0,\qquad
\{\widetilde{{\cal H}},\widetilde{\Omega}\}=0,
\eeq
and the boundary
conditions
\beq
\widetilde{\Omega}\overleftarrow{\pa}_{\!\!C^{\alpha}}\big|_{C=0}=
\widetilde{T}_{\alpha},\quad \widetilde{{\cal
H}}=\widetilde{H}_0+\cdots,
\eeq
where $"\cdots"$ in the last
expression correspond to power-series expansions in the ghost
variables. We can search for the solutions to the basic equations
(\ref{baseq}) in the form of series in deformation
parameter $g$,
\beq
\widetilde{\Omega}=\Omega+g\Omega_{(1)}+g^2\Omega_{(2)}+\cdots
,\quad \widetilde{{\cal H}}={\cal H}+g{\cal H}_{(1)}+ g^2{\cal
H}_{(2)}+\cdots ,
\eeq
 where $\Omega$   and ${\cal H}$  are the
BRST-BFV charge and the Hamiltonian respectively for initial system.
They satisfy the basic equations (\ref{beOH}). As consequence, such
a deformation procedure is described by the following infinite
system of equations
\beq
\label{sq1}
&&2\{\Omega,
\Omega_{(1)}\}=0,\quad 2\{\Omega,\Omega_{(2)}\}+\{\Omega_{(1)},\Omega_{(1)}\}=0,\\
\label{sq2} &&\{{\cal H}, \Omega_{(1)}\}+\{{\cal H}_{(1)},
\Omega\}=0,\quad \{{\cal H}, \Omega_{(2)}\}+\{{\cal
H}_{(1)},\Omega_{(1)}\}+ \{{\cal H}_{(2)},\Omega\}=0, \eeq and so
on. Note that solutions to the equations (\ref{sq1}), (\ref{sq2})
for some simple dynamical systems (in lower orders) are searched
with the help of the cohomological methods (see e.g. \cite{BMS},
\cite{BCCSS}, \cite{BCSSI}, \cite{Dai}). It is clear that it is
difficult to expect to find a general solution to the infinite
system of equations (\ref{baseq}) in a compact and closed form. The
above deformation procedure is described as an infinite series and
therefore leads to the obvious problem of its convergence. Besides,
it is unclear from the very beginning that the sum of the series
really exists and whether it is local or no. In particular, it is
not difficult to construct an example of non-local functional whose
each term of expansion in power series will be local.

In our paper, we propose a new  way for describing the consistent
deformations of dynamical systems with constraints, which does not
require expansion into series at all. This new way is based on the
fundamental property of the basic equations (\ref{baseq}) and
(\ref{beOH}) being invariant under (super)canonical transformations
in the extended phase space.

\section{Solutions to deformation problem}

In this section, we describe a general approach to the solution of the deformation problem.
Let $(Q^A,P_A)=(q^i,p_i;C^{\alpha}, {\cal P}_{\alpha})$ are the initial
minimal extended phase space
variables satisfying the commutation relations in terms of (super)Poisson brackets
\beq
\{q^i,p_j\}=\delta^i_j,\quad \{C^{\alpha}, {\cal P}_{\beta}\}=\delta^{\alpha}_{\beta},
\eeq

In this phase space, one considers the (super)canonical transformation of the form
\beq
Q^{'A}=\overrightarrow{\pa}_{P^{'}_A}Y(Q,P^{'}),\quad
P_A=Y(Q,P^{'})\overleftarrow{\pa}_{Q^A},
\eeq
where $Y(Q,P^{'}), \;(\varepsilon(Y(Q,P^{'})=0)$ is the corresponding generating
function of the transformation.
Then the canonically transformed BRST-BFV charge
\beq
{\Omega}^{'}={\Omega}^{'}(Q,P)={\Omega}(Q^{'}(Q,P),P^{'}(Q,P)),
\eeq
and the Hamiltonian
\beq
{\cal H}^{'}={\cal H}^{'}(Q,P)={\cal H}(Q^{'}(Q,P), P^{'}(Q,P))
\eeq
automatically satisfy the same basic equations
\beq
\label{becan}
\{{\Omega}^{'},{\Omega}^{'}\}=0,\qquad
\{{\cal H}^{'},{\Omega}^{'}\}=0
\eeq
as well as $\Omega$ and ${\cal H}$ (\ref{beOH}).

We will search for the generating function of the canonical transformation in the form
\beq
Y(Q,P^{'})=P^{'}_AQ^A+X(Q,P^{'}),
\eeq
where first term means the identical transformation and  $X(Q,P^{'})$
is some unknown yet function.
It yields
\beq
Q^{'A}=Q^A+\overrightarrow{\pa}_{P^{'}_A}X(Q,P^{'}),\quad
P_A=P^{'}_A+X(Q,P^{'})\overleftarrow{\pa}_{Q^A}
\eeq

For further consideration, we choose the function $X(Q,P^{'})$ as follows
\beq
X(Q,P^{'})=p^{'}_ih^i(q)
\eeq
with some new function $h^i(q).$ Then, it is easy to see that such
a canonical transformation non-trivially
affects only the variables $q^{i},p_{i}$ of the initial phase space.
The corresponding transformation
has the form
\beq
q^{'i}=q^i+h^i(q),\quad p^{'}_i=p_j(M^{-1}(q))^j_{\;i},
\label{transformations}
\eeq
where the matrix $(M^{-1}(q))^j_{\;i}$ is inverse to the following one
\beq
M^i_{\;j}(q)=\delta^i_{\;j}+h^i(q)\overleftarrow{\pa}_{q^j}.
\label{M}
\eeq

As a result, one gets for the transformed BRST-BFV charge and Hamiltonian
the following expressions
\beq
&&\widetilde{\Omega}=\widetilde{\Omega}(Q,P)=
\Omega(q+h(q), pM^{-1}(q), C,{\cal P}),\\
&&\widetilde{{\cal H}}=\widetilde{{\cal H}}(Q,P)= {\cal H}(q+h(q),
pM^{-1}(q), C,{\cal P}), \label{transformed}
\eeq
which satisfy the
equations
\beq
\{\widetilde{\Omega}, \widetilde{\Omega}\}=0,\qquad
\{\widetilde{{\cal H}},\widetilde{\Omega}\}=0
\eeq and the boundary
conditions
\beq
\widetilde{\Omega}\overleftarrow{\pa}_{\!\!C^{\alpha}}\big|_{C=0}=
\widetilde{T}_{\alpha},\qquad
\widetilde{{\cal H}}=\widetilde{H}_0+\cdots ,
\eeq
where $\cdots$
means the corrected terms stipulated by the transformations
(\ref{transformed}). Here we have used the notations
\beq
\label{37}
&&\widetilde{H}_0=\widetilde{H}_0(q,p)=
H_0(q+h(q), pM^{-1}(q)),\\
\label{38}
&&\widetilde{T}_{\alpha}=\widetilde{T}_{\alpha}(q,p)=
T_{\alpha}(q+h(q), pM^{-1}(q)).
\eeq
In its turn, the functions $\widetilde{H}_0$ and $\widetilde{T}_{\alpha}$
satisfy the involution equations
\beq
\label{39}
\{\widetilde{H}_0, \widetilde{T}_{\alpha}\}=
\widetilde{T}_{\beta}\widetilde{V}^{\beta}_{\;\!\alpha},\quad
\{\widetilde{T}_{\alpha},\widetilde{T}_{\beta}\}=
\widetilde{T}_{\gamma}\widetilde{U}^{\gamma}_{\;\;\alpha\beta},
\eeq
where
\beq
\label{40}
&&\widetilde{V}^{\beta}_{\;\!\alpha}=
\widetilde{V}^{\beta}_{\;\!\alpha}(q,p)=
V^{\beta}_{\;\!\alpha}(q+h(q), pM^{-1}(q)),\\
\label{41} &&\widetilde{U}^{\gamma}_{\;\;\alpha\beta}=
\widetilde{U}^{\gamma}_{\;\;\alpha\beta}(q,p)=
U^{\gamma}_{\;\;\alpha\beta}(q+h(q), pM^{-1}(q)) .
\eeq
From
Eqs. (\ref{37}), (\ref{38}), (\ref{40}), (\ref{41}) it follows that
all quantities of deformed theory satisfying the equations
(\ref{39}) are obtained from initial ones with the help of change of
variables (\ref{transformations}). Thus, we see that the deformed
theory is described by the first-class constraints
$\widetilde{T}_{\alpha}$, Hamiltonian $\widetilde{H}_0$ and
structure functions $\widetilde{V}^{\beta}_{\;\!\alpha}, \,
\widetilde{U}^{\gamma}_{\;\;\alpha\beta}$ which are constructed in
the explicit form on the base of constraints ${T}_{\alpha}$,
Hamiltonian ${H}_0$ and the structure functions
$V^{\beta}_{\;\!\alpha}, \, U^{\gamma}_{\;\;\alpha\beta}$ of the
initial theory with help of an arbitrary function $h(q)$. It is
important to point out that no a priori restrictions of the function
$h(q)$ are imposed \footnote{Of course it is assumed that this
function is differentiable and transforms as coordinates.}.
It is also worth noting that the deformed BRST-BFV charge
and Hamiltonian have the structures which are the same as in (\ref{sOH}) with natural
replacement of the initial quantities by the deformed ones.

Thus, we have constructed a special type of canonical transformation
which maps an initial dynamical system with first-class constraints
into the deformed dynamical system with first-class constraints. The
new dynamical system fully satisfies all the properties of
deformation procedure mentioned in the previous section. We have
proved that the deformation procedure in the generalized canonical
formalism can be described by only one generating function $h(q)$,
which depends on the coordinates $q^i$ of the initial phase space.
Moreover, it follows from the above relations that the function
$h(q)$ indeed plays the role of the deformation parameter. We see
that the description of the deformation procedure in the canonical
formalism is similar to the description in the Lagrangian formalism
\cite{BL-1}, \cite{BL-2}, \cite{L}, where also only one arbitrary
function is needed to encode the deformation. Compared to the
cohomological approach to the problem of deformation, our method
allows us to give general answers to questions concerning the
structure of the deformation. First, all the arbitrariness in the
deformation procedure is completely controlled by one generating
function $h(q)$ of initial coordinates. Secondly, the result is
completely obtained in a closed form and does not require expansion
into series. In fact, this result can be interpreted as the sum of a
series in the cohomological approach. Thirdly, in obtaining the
general result, no assumptions are made about the locality or
non-locality of the deformation structure.
Therefore, the deformation must be nonlocal in the general case if we
are interested in construction of classically non-equivalent theories.

\section{Deformation of Abelian gauge theory into non-Abelian}
\noindent
In this section, we will show that within the framework of the
canonical formalism one can construct such a deformation that
transforms an Abelian gauge theory into some non-local non-Abelian  gauge theory with
a local sector corresponding to Yang-Mills theory.
The structure of a gauge theory in the canonical formalism is defined
by a system of first-class constraints and the Hamiltonian in involution with
these constraints. Therefore, we will show that there exists a non-local
deformation of the fields
and conjugate momenta, which transforms, in local sector, the constraints and
the Hamiltonian of the
Abelian gauge theory into the constraints and the Hamiltonian of the
Yang-Mills theory.
The concrete choice of the function $h^{i}$ for appropriate deformation
is based on the relations
(\ref{transformations}), (\ref{M}).

Let us consider the $N$ copies of the Abelian gauge theory with
fields $A^{a}{}_{\mu}, \, a=1,2,\ldots, N; \, \mu=(0,i); i=1,2,3.$ In the canonical
formalism, such a theory is described by the coordinates
$A^{'a}_{0}(\vec{x}),\, A^{'a}_{i}(\vec{x})$ and the conjugate momenta
$p^{'a}_{0}(\vec{x}),\, p^{'a}_{i}(\vec{x})$ \footnote{In this section
$i,j,k$ mean three-dimensional spatial indices.} obeying the first-class constraints
\beq
p^{'a}_{0}=0, \, \, \partial_{i}p^{'a}_{i}=0.
\label{abelain constraints}
\eeq
The corresponding Hamiltonian has the form
\beq
H_{Abelian}=\int d^3x \big(\frac{1}{2}p^{'a}_{i}p^{'a}_{i} -A^{'a}_{0}\partial_{i}p^{'a}_{i}+
\frac{1}{4}F^{'a}_{ij}F^{'a}_{ij}\big),
\label{H1}
\eeq
where $F^{'a}_{ij}=\partial_{i}A^{'a}_{j}-\partial_{j}A^{'a}_{i}.$

On the other hand, one considers the Yang-Mills theory with the fields $A^{a}{}_{\mu}$
and the structure constants $f^{abc}$. In the canonical formalism this
theory is described by the coordinates $A^{a}_{0}(\vec{x}),\, A^{a}_{i}(\vec{x})$ and the
conjugate momenta $p^{a}_{0}(\vec{x}),\, p^{a}_{i}(\vec{x})$ obeying the first-class
constraints
\beq
p^{a}_{0}=0, \, \, D^{ab}_{i}p^{b}_{i}=0,
\label{nonabelain constraints}
\eeq
where
\beq
D^{ab}_{i}p^{b}_{i}=\partial_{i}p^{a}_{i}+gf^{acb}A^{c}_{i}p^{b}_{i}
\label{D}
\eeq
and $g$ is a coupling constant. The corresponding Hamiltonian has the form
\beq
H_{non-Abelian}=\int d^3x \big(\frac{1}{2}p^{a}_{i}p^{a}_{i}
-A^{a}_{0}D^{ab}_{i}p^{b}_{i}+ \frac{1}{4}G^{a}_{ij}G^{a}_{ij}\big),
\label{H2}
\eeq
where $G^{a}_{ij}=\partial_{i}A^{a}_{j}-\partial_{j}A^{a}_{i} + gf^{abc}A^{b}_{i}A^{c}_{j}.$
Derivation of the constraints and the Hamiltonians for Abelian and non-Abelian
gauge theories see e.g.
in \cite{GT}, \cite{RR}.

We will explore the deformation (\ref{transformations}) of the form
\beq
&&A^{'a}_{0} = A^{a}_{0}, \, \quad p^{'a}_{0}= p^{a}_{0},\\
&&A^{'a}_{i}(\vec{x})={A}^{a}_{i}(\vec{x})+h^{a}_{i}(\vec{x}), \,\quad
p^{'a}_{i}({\vec{x}})= \int d^3y \,p^{b}_{j}({\vec{y}})
(M^{-1})^{ba}_{ji}({\vec{y}},{\vec{x}}),
\label{transformations1}
\eeq
where $h^{a}_{i}$ depends on $A^{a}_{i}({\vec{x}})$ and, accordingly to (\ref{M}),
the matrix $(M)^{ab}_{ij}(\vec{x},\vec{y})$ looks like
\beq
(M)^{ab}_{ij}(\vec{x},\vec{y}) = \delta^{ab}\delta_{ij}\delta(\vec{x}-\vec{y}) +
\frac{\delta h^{a}_{i}(\vec{x})}{\delta A^{b}_{j}(\vec{y})}.
\label{M1}
\eeq
with $\vec{x}=(x^{i}; i=1,2,3).$

Let us consider the non-local function $h^{a}_{i}(\vec{x})$ in
(\ref{transformations1}) as follows
\beq
h^{b}_{j} =
gf^{bcd}\frac{1}{\Delta}\Big(\partial_{k}(A^{c}_{k}A^{d}_{j})+
\frac{1}{4}f^{cmn}A^{d}_{k}A^{m}_{k}A^{n}_{j}\Big),
\label{h}
\eeq
where $\frac{1}{\Delta}$ is the inverse
Laplacian.  Here and further, the spatial indices $\vec{x}, \vec{y}$
suppressed. It is evident that the obtained deformed theory should be non-local.
Our goal is to explore the possibility that such a nonlocal theory has a local sector and to
describe this local sector.  Taking into account the expression
(\ref{M1}), we get for inverse matrix
\beq
(M^{-1})^{ba}_{ji} = \delta^{ab}\delta_{ij}
- \frac{\delta h^{b}_{j}}{\delta A^{a}_{i}}+\ldots,
\label{inverseM}
\eeq
where $\ldots$ mean the terms of higher order in coupling
constant. The relation (\ref{inverseM}) yields
\beq
\label{inverseM1}
&&(M^{-1})^{ba}_{ji} = \delta^{ab}\delta_{ij} +
gf^{bad}\partial_{i}\frac{1}{\Delta}A^{d}_{j} +\\
\nonumber
&&\qquad\qquad\quad +\; \textit{the
non-local terms unessential for further consideration}.
\eeq
The result (\ref{inverseM1}) allows to write
the deformation of the momenta $p^{'a}_{i}({\vec{x}})$ in
(\ref{transformations1}) as follows
\beq
\label{trasformations2}
&&p^{'a}_{i}=
p^{a}_{i}+gf^{bad}p^{b}_{j}\partial_{i} \Delta^{-1}A^{d}_{j}+\\
\nonumber
&&\qquad\;\; +\;
\textit{the non-local terms unessential for further consideration}.
\eeq
Substituting the expression
(\ref{trasformations2}) into Abelian constraint
$\partial_{i}p^{'a}_{i}$ one obtains \beq
\partial_{i}p^{'a}_{i} = D^{ab}_ip^{b}_{i}(\vec{x}) + \textit{non-local terms}.
\label{nonabelian constraints1}
\eeq
The local term in (\ref{nonabelian constraints1}) corresponds
to  constraint in the non-Abelian theory
(\ref{nonabelain constraints}). Another constraint $p^{a}_{0}=0$
is fulfilled automatically since the
$p^{a}_{0}$ does not transform.

Now we turn to the Hamiltonian $H_{Abelian}$ (\ref{H1}) and perform the transformations
(\ref{transformations1}) with function $h^{a}_{i}(\vec{x})$ (\ref{h}). It is evident that
$p^{'a}_{i}p^{'a}_{i} = p^{a}_{i}p^{a}_{i} +\textit{non-local terms}$ and also
$A^{'a}_{0}\partial_{i}p^{'a}_{i} = A^{a}_{0}D^{ab}_{i}p^{b}_{i}+\textit{non-local terms}$,
where we have used the relation (\ref{nonabelian constraints1}). The expression
$F^{a}_{ij}F^{a}_{ij}$ in the Hamiltonian $H_{Abelian}$ (\ref{H1})
is transformed by the same way as the term $F^{\mu\nu}F_{\mu\nu}$
in section 5 of ref. \cite{BL-1}
with replacement $\mu,\nu, \Box$ by $i,j, \Delta$ respectively.
That is, $F^{a}_{ij}F^{a}_{ij}$ takes
the form $G^{a}_{ij}G^{a}_{ij} + \textit{non-local terms}$.
Therefore, the Hamiltonian $H_{Abelian}$
(\ref{H1}) transforms into the Hamiltonian $H_{non-Abelian}$ (\ref{H2})
up to non-local terms.

Thus, we see that the deformation (\ref{transformations1})
with non-local function $h^{a}_{i}$ (\ref{h})
maps Abelian gauge theory into some non-local non-Abelian gauge theory that contains
a local sector coinciding with Yang-Mills theory. Let us stress once more that
due to the non-locality of deformation the obtained non-Abelian local theory is not
classically equivalent to original Abelian local theory.
It is a special local sector of a complicated non-local theory.

\section{Summary}
\noindent In the present paper we have studied the deformation
problem for an arbitrary dynamical system with irreducible
first-class constraints in phase space and proposed a general method
of solution to this problem. Our approach is based on the
BFV-formalism, where the dynamical system is formulated in the
extended phase space and its gauge structure is described by the
BRST-BFV charge $\Omega$ and the generalized Hamiltonian ${\cal H}$
satisfying the equations (\ref{beOH}) in terms of the Poisson
superbrackets. Structure of the initial dynamical system is encoded
by the boundary conditions (\ref{bcOH}). The generating equations
(\ref{beOH}) are invariant under (super)canonical transformations of
the phase space variables (\ref{becan}).

We describe the deformation procedure in terms of special minimal
(super)canonical transformation and construct this transformation in
explicit and closed form. It is proved that such a transformation is
defined by a single generating function $h(q)$ which depends only on
coordinates of the initial phase space. Deformations of the initial
Hamiltonian $H_0$ and the constraints $T_{\alpha}$ are described as
change of variables in the arguments of these quantities, namely the
shift of coordinates, $q^i+h^i(q)$, and the rotation of momenta,
$p_j(M^{-1}(q))^j_{\;i}$, with the matrix inverse to matrix
$M^i_{\;j}(q)=\delta^i_{\;j}+h^i(q)\overleftarrow{\pa}_{\!q^j}$. The
function $h(q)$ is completely arbitrary, there is no a priori
requirements on this function, in particular, it can be non-local.
As a result, we find a solution to the deformation problem in closed
and constructive form. The choice of the generating function just as
$h(q)$ corresponds in fact to fixing the arbitrariness existing in
the Hamiltonian gauge algebra for structure functions of involution
equations \cite{BT}.

To illustrate the possibilities of our method, we have applied the
developed deformation procedure to system of several copies of an
Abelian gauge theory formulated in the canonical formalism. We have
constructed an appropriate function $h(q)$ mapping such a theory
into a non-local non-Abelian gauge theory that has a local sector
coinciding with the Yang-Mills theory in canonical formalism.

One can expect that the considered deformation procedure within the
framework of BFV-formalism can be applied to construct the new gauge
theories on the based of known gauge theories. As we have already
mentioned, the deformation procedure is in general non-local. We
believe that the most interesting aspect of applying our deformation
procedure to concrete theories is the possibility of that a local
sector exists in {\bf the resultant} non-local theory, after
deformation. It means that the non-local deformed action and
non-local deformed gauge transformation may in principle have
completely local sector which define a new local gauge theory
obtained on the base of the given local gauge theory. These aspects
deserve a special study in case of each concrete theory.

It is worth mentioning out that just the BFV-formalism is used in
the modern higher spin theory to derive the interacting vertices for
higher spin fields (see e.g. the recent papers \cite{BKTW},
\cite{BR}, \cite{BKS} and the references therein). One can expect that the
developed deformation procedure could allow to generate the
interacting vertices by the deformation of free higher spin
theories. Such a possibility in principle was demonstrated in \cite{BL-1} within
the framework of the Lagrangian deformation procedure\footnote{See
the new applications of this approach in higher spin field theory in
\cite{L1}, \cite{L2}.}. We plan to consider the applications of our canonical
deformation procedure in higher spin field theory in forthcoming
papers.  Another interesting issue is to derive
a complete general relativity by
suitable deformation of a linearized massless spin 2 theory.
In principle, such a deformation was constructed by summing infinite
series in \cite{OP}, \cite{BD}. We believe that our method allows us to carry out
a similar study without using series expansions.

\section*{Acknowledgments}
\noindent The work is supported by the Ministry of Education of the
Russian Federation, project QZOY-2023-0003.

\section*{Data Availability Statement}

Data Availability Statement: No Data is associated with the
manuscript.

\end{document}